\documentclass[aps,pre,preprint,superscriptaddress,showpacs]{revtex4-1}

\usepackage{graphicx}
\usepackage{dcolumn}
\usepackage{bm}

\begin{document}


\title{Modeling of photon and pair production due to quantum electrodynamics effects in particle-in-cell simulation}

\author{W.-M. Wang}
\affiliation{Beijing National Laboratory for Condensed Matter
Physics, Institute of Physics, CAS, Beijing 100190, China}
\author{Z.-M. Sheng}
\affiliation{Department of Physics, SUPA, Strathclyde University,
Rottenrow 107, G4 0NG Glasgow, United Kingdom} \affiliation{Key
Laboratory for Laser Plasmas (MoE) and Department of Physics and
Astronomy, Shanghai Jiao Tong University, Shanghai 200240, China}
\author{P. Gibbon}
\affiliation{Forschungzentrum Juelich GmbH, Institute for Advanced
Simulation, Juelich Supercomputing Centre, D-52425 Juelich, Germany}
\author{Y.-T. Li}
\affiliation{Beijing National Laboratory for Condensed Matter
Physics, Institute of Physics, CAS, Beijing 100190, China}

\date{\today}

\begin{abstract}
We develop the particle-in-cell (PIC) code KLAPS to include the
photon generation via the Compton scattering and electron-positron
creation via the Breit-Wheeler process due to quantum
electrodynamics (QED) effects. We compare two sets of existing
formulas for the photon generation and different Monte Carlo
algorithms. Then we benchmark the PIC simulation results.
\end{abstract}


\maketitle

With the development of ultraintense laser technology, 10-PW-class
laser pulses will be available soon worldwide. A few of 100-PW-class
laser systems are also under construction, e.g., the ELI system in
Europ \cite{ELI}, the OMEGA EP-OPAL laser system in USA
\cite{OMEGA_EP_OPAL}, etc. The focused laser intensity will exceed
$10^{23}\rm W cm^{-2}$ and even reach $10^{25}\rm W cm^{-2}$. Under
irradiation of so high intensity laser pulses, electrons will be
quickly accelerated to have energy at the GeV scale. Interaction of
the high-energy electrons with the laser pulse, a large number of
$\gamma-$photons will be generated via the Compton scattering since
the QED parameter \cite{Erber,Kirk,Elkina} of $\chi_e\simeq \gamma
F_\perp/(eE_S)$ can exceed 1, where $\gamma$ is the electron lorentz
factor, $E_S=1.32\times10^{18}V/m$ is the Schwinger field
\cite{Schwinger1,Schwinger2} and $F_\perp$ is the transverse
component of the Lorentz force. If the generated photons have high
enough energy to make the QED parameter of photons $\chi_{ph}\simeq
(\hbar\omega/m_ec^2) F_\perp/(eE_S)$ approaching 1,
electron-positron pairs will be created via a Breit-Wheeler process
\cite{Erber,Kirk,Elkina}. Therefore, it is necessary to include such
pair creation and photon generation in the simulation for the newly
developed laser pulse interaction. In this paper, we develop our PIC
code KLAPS \cite{KLAPS} to include such QED processes.

Under quasi-stationarity and weak-field approximations
\cite{Erber,Kirk,Elkina}, two different sets of formulas are taken
respectively to calculate the photon and pair generation rate.
Considering that a positron with the same velocity as an electron
has the same photon generation rate with the electron, we just give
the expression with respect to electrons in the following. One
formula is given by \cite{Elkina,Nerush}:
\begin{eqnarray}
\frac{d W_{rad}}{d\xi}=\frac{\alpha m_e c^2}{\sqrt{3} \pi \hbar
\gamma_e} [(1-\xi +\frac{1}{1-\xi}) K_{2/3}(\delta) -
\int_{\delta}^{\infty} K_{1/3}(s)ds ],
\end{eqnarray}
where $\xi=\varepsilon_{ph}/\varepsilon_e$, $\varepsilon_{ph}=m_e
c^2 \gamma_{ph}$ is the generated photon energy, $\varepsilon_e=m_e
c^2 \gamma_e$ is the electron energy, $\delta=2\xi /[3(1-\xi)
\chi_e]$, $\alpha\simeq 1/137$, $K_{\nu}$ is a modified Bessel
function. The QED parameters $\chi_e$, $\chi_{ph}$ with respect to
the electron and photon are defined as:
\begin{eqnarray}
\chi_e= \frac{\gamma_e}{E_S} \sqrt{(\mathbf{E}+ \mathbf{v_e} \times
\mathbf{B})^2 - (\mathbf{v_e} \cdot \mathbf{E})^2},
\end{eqnarray}
and
\begin{eqnarray}
\chi_{ph}= \frac{\gamma_{ph}}{E_S} \sqrt{(\mathbf{E}+
\mathbf{v_{ph}} \times \mathbf{B})^2 - (\mathbf{v_{ph}} \cdot
\mathbf{E})^2},
\end{eqnarray}
where $E_S=1.32\times10^{18}V/m$ is the Schwinger field
\cite{Schwinger1,Schwinger2}, $\mathbf{v_e}$ and $\mathbf{v_{ph}}$
normalized by $c$ are velocities of the electron and photon,
$\mathbf{E}$ and $\mathbf{B}$ normalized by $m_ec\omega_0/e$ are the
electric and magnetic fields experienced by the electron and photon.

The other formula for the photon generation rate is given by
\cite{Erber,Kirk}:
\begin{eqnarray}
\frac{d W_{rad}}{d\xi}=\frac{\alpha m_e c^2 \xi}{3\pi^2 \hbar
\gamma_e \chi_{e}} [\sum_{i=1}^3 F_i(\xi)J_i(\sigma)],
\end{eqnarray}
where $\sigma=\frac{\xi}{3\chi_e(1-\xi)}$,
$F_1(\xi)=1+(1-\xi)^{-2}$, $F_2(\xi)=2(1-\xi)^{-1}$, $F_3(\xi)=\xi^2
(1-\xi)^{-2}$,

$J_1(\sigma)=\frac{1}{3\sigma^2} \int_{\sigma}^\infty du
\frac{u}{\sqrt{(u/\sigma)^{2/3}-1}} K_{2/3}^2(u)$,

$J_2(\sigma)=\frac{1}{3\sigma} \int_{\sigma}^\infty du
(u/\sigma)^{1/3} \sqrt{(u/\sigma)^{2/3}-1} K_{1/3}^2(u)$,

$J_3(\sigma)=\frac{1}{3\sigma^2} \int_{\sigma}^\infty du
\frac{u}{\sqrt{(u/\sigma)^{2/3}-1}} K_{1/3}^2(u)$.

In the classic limit with $\hbar\rightarrow0$, the photon generation
rate is reduced to
\begin{eqnarray}
\frac{d W_{rad}}{d\xi}=\frac{\sqrt{3}\alpha m_e c^2 \chi_{e}}{2\pi
\hbar \gamma_e \xi} \zeta\int_\zeta^\infty du K_{5/3}(u),
\end{eqnarray}
where $\zeta=2\xi/(3\chi_e^2)$.

\begin{figure}[htbp]
\includegraphics[width=6in]{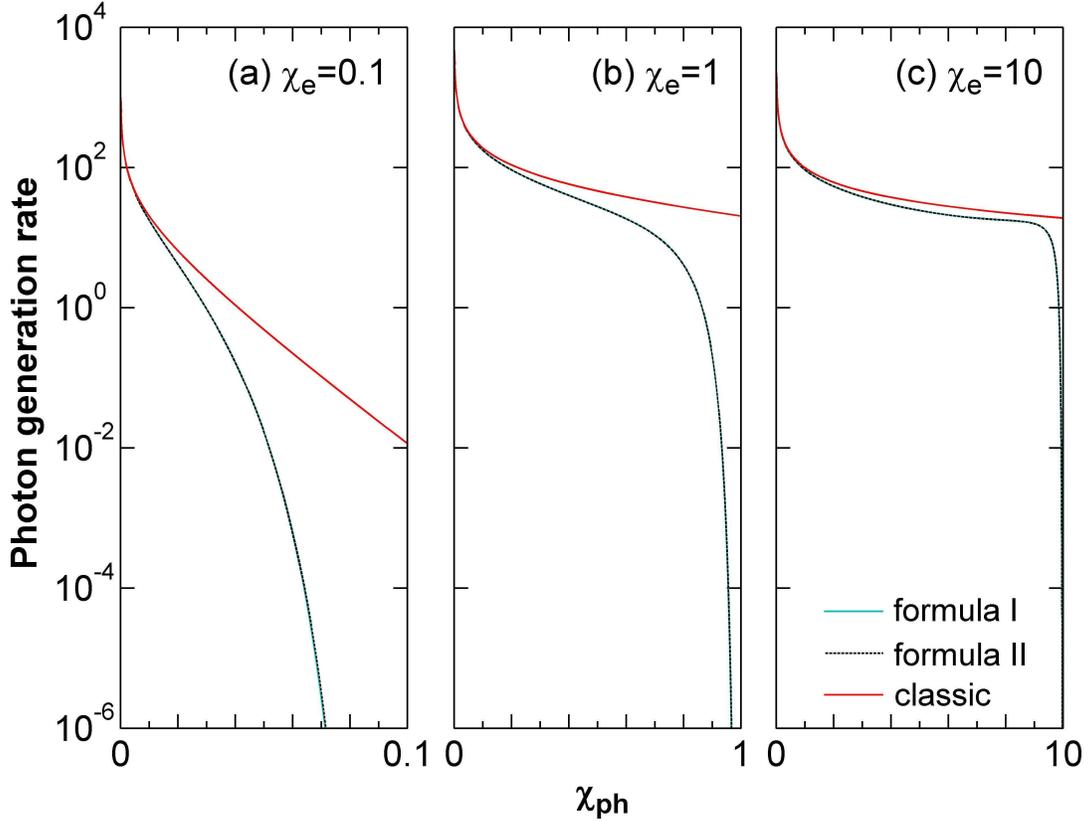}
\caption{\label{fig:epsart} Photon generation rates calculated by
formula I [Eq. (1)], formula II [Eq. (4)], and the classic formula
[Eq. (5)], respectively. Plots (a), (b), and (c) correspond to
different $\chi_{e}$. }
\end{figure}

We numerically calculate Eqs. (1), (4), and (5), which are denoted
by ``formula I", ``formula II", and ``classic", respectively, in
Fig. 1. We take a B field with the strength of $b_0$ transverse to
the electron motion plane. In Fig. 1(a), $b_0=200 m_ec\omega_0/e$,
$\gamma_e=206$ and $\chi_e=0.1$; in Fig. 1(b) $b_0=2000
m_ec\omega_0/e$, $\gamma_e=206$ and $\chi_e=1$; and in Fig. 1(c)
$b_0=2000 m_ec\omega_0/e$, $\gamma_e=2060$ and $\chi_e=10$. One can
see that the formula I and II are nearly the same with different
$\chi_e$. The classic formula overestimates the rate at the
high-energy photon range, as expected. Therefore, one can use either
the formula I or the formula II. In the following part and in our
simulation we adopt the formula I [Eq. (1)]. Then, we take the pair
generation rate \cite{Elkina,Nerush}, which have the similar form
with Eq. (1). It is given by:
\begin{eqnarray}
\frac{d W_{pair}}{d\xi}=\frac{\alpha m_e c^2}{\sqrt{3} \pi \hbar
\gamma_{ph}} [(\frac{1}{\xi} +\frac{1}{1-\xi}-2) K_{2/3}(\delta) -
\int_{\delta}^{\infty} K_{1/3}(s)ds ],
\end{eqnarray}
where $\delta=2/[3\xi (1-\xi) \chi_{ph}]$,
$\xi=\varepsilon_e/\varepsilon_{ph}=\gamma_e/\gamma_{ph}$, and the
energy of the created positron is
$\varepsilon_p=\varepsilon_{ph}-\varepsilon_e$.

\begin{figure}[htbp]
\includegraphics[width=3.6in]{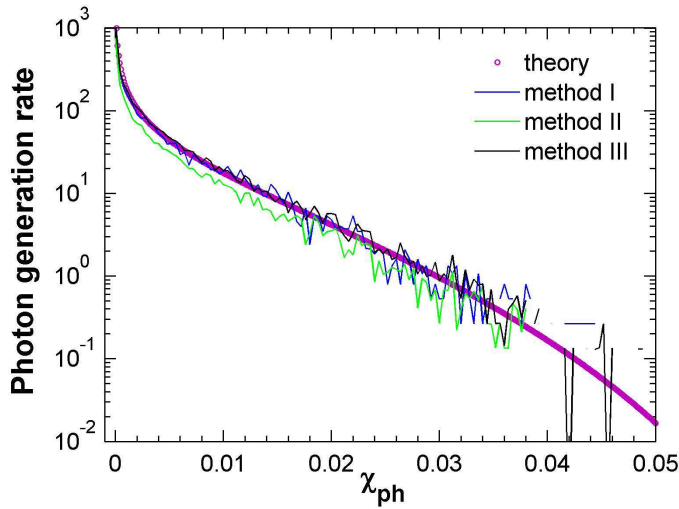}
\caption{\label{fig:epsart} Average rate of photon generation
obtained from PIC simulations, where the theoretic values are given
by Eq. (1). Three different event generator for photon generation
are adopted.}
\end{figure}

In Figs. 2-4, we benchmark the Monte Carlo simulations by our
QED-PIC code against the numerical calculations of Eqs. (1) and (6),
respectively. In the simulations, we take $128 \times 192$ cells (in
$x\times y$) and 1 electron with $\gamma_e=206$ per cell. The
simulation is run 480 time steps. The obtained average rate of
photon generation is shown in Fig. 2. Three methods are respectively
adopted for the event generator. Method I: firstly the total
generation rate $W_{rad}$ is computed; if $W_{rad} Dt>r_1$, a photon
will be generated, where $r_1$ ($0<r_1<1$) is a
uniformly-distributed random number; the photon energy with
$\varepsilon_{ph}=\xi^0 \times \varepsilon_{e}$ is obtained through
\begin{eqnarray}
\int_{\xi_{min}}^{\xi^0} \frac{d W_{rad}}{d\xi}=r_2W_{rad},
\end{eqnarray}
where $r_2$ ($0<r_2<1$) is another uniformly-distributed random
number, independent of $r_1$. Here, the lower limit of integration
$\xi_{min}$ is set to avoid the infrared singularity, where $
\xi_{min} \times \varepsilon_{e}=2m_ec^2$. Method II: firstly a
uniformly-distributed random number $r_3$ is taken; then a
cumulative probability $P_{cum}$ is calculated by
$P_{cum}=P_{cum}+W_{rad} Dt$; if $1-\exp(-P_{cum})>r_3$, a photon
will be generated and the photon has an energy of
$\varepsilon_{ph}=\xi^0 \times \varepsilon_{e}$, where $\xi^0$ is
obtained through Eq. (7). Method III is similar to the method II,
except that the condition of a photon generation is changed to
$P_{cum}>r_3$. One can see the three methods shows equivalent, as
seen in Fig. 2.

\begin{figure}[htbp]
\includegraphics[width=6in]{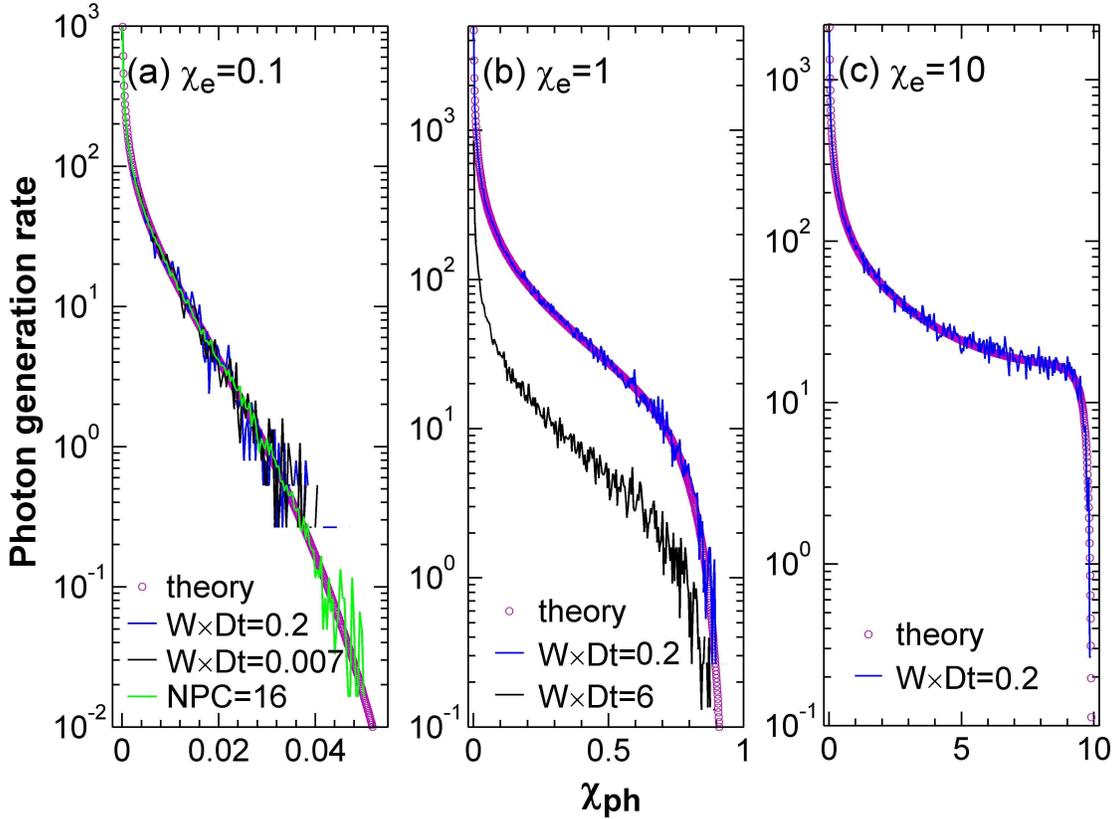}
\caption{\label{fig:epsart} Photon generation rates are obtained
from the theory given by Eq. (1), and simulations with different
time resolution Dt and different number of particles per cell. In
(a)-(c), different $\chi_{e}$ is taken, respectively.}
\end{figure}

\begin{figure}[htbp]
\includegraphics[width=6in]{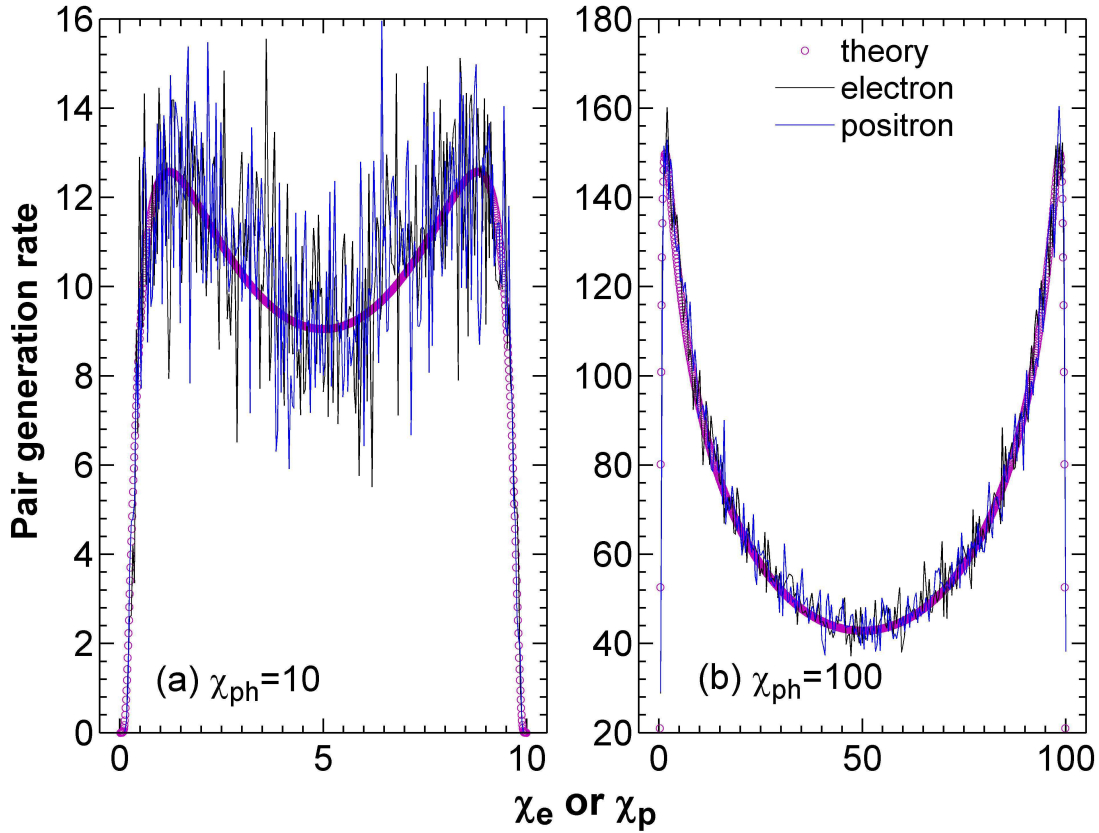}
\caption{\label{fig:epsart} Electron-positron pair generation rates
are obtained from the theory given by Eq. (1), and simulations with
Dt=$0.2/W$, where different $\chi_{ph}$ is taken in (a) and (b). }
\end{figure}

In the following simulations, we just take first even generator for
both the photons and pairs. Figures 3 and 4 shows the comparison of
the photon and pair generation rates given by our PIC simulations
against the numerical calculations of Eqs. (1) and (6),
respectively. One can see that the two results are in good agreement
with different $\chi_{ph}$. We have taken the time step as
$Dt=0.2/W$, $W$ is the total rate of photon or pair generation. When
the time step is increased to $Dt=6/W$ in Fig. 3(b), the simulation
is very different from the theoretical values. When a small enough
time step $Dt=0.007/W$ is also taken in Fig. 3(a), the simulation
result is nearly the same with the one with $Dt=0.2/W$.

Then, we take an adjustable time step such as when the $W\times
Dt>0.2$, the particle generator is automatically separated $N$
steps/circles to meet $W\times Dt/N \leq 0.2$.

\begin{figure}[htbp]
\includegraphics[width=6in]{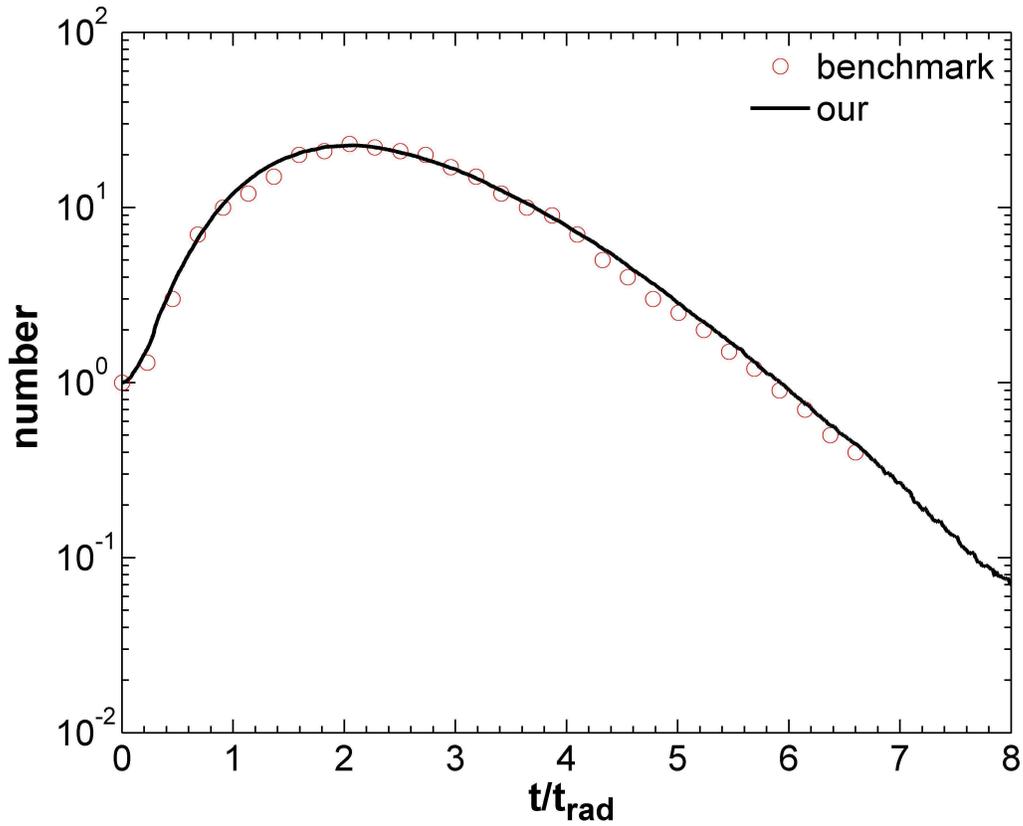}
\caption{\label{fig:epsart} Number of pairs with energy above 100
MeV are created from a cascade. The benchmark data are obtained from
the QED-PIC result in Ref. \cite{Elkina}. }
\end{figure}

Finally, we benchmark our code against the QED-PIC simulation result
\cite{Elkina} on a cascade development from a single electron with
$\gamma=2\times 10^5$ initially under a static external magnetic
field of $0.2 E_S$ perpendicularly to the electron motion plane. We
counter the created pairs with energy above 100 MeV, as shown in
Fig. 5. It is shown that our results agree with the result in Ref.
\cite{Elkina}. Here, $t_{rad}=1.16\times10^{-16}$ is taken as a
characteristic radiation time. Our results are averaged over 4000
simulation runs.

In summary, we have developed our code KLAPS to include QED
processes via Monte Carlo methods. This QED-PIC allow us to
investigate QED-dominant laser plasma interaction.

\begin{acknowledgments}
This work was supported by the National Basic Research Program of
China (Grants No. 2013CBA01500) and NSFC (Grants No. 11375261, No.
11105217, No. 11121504, and No. 113111048).
\end{acknowledgments}

%

\end{document}